\documentstyle[12pt]{article}
\newcommand{\bo}{B^0}
\newcommand{\bob}{\bar{B^0}}
\newcommand{\bq}{B_q}
\newcommand{\bqb}{\bar{B_q}}

\newcommand{\bs}{B_s}
\newcommand{\bsb}{\bar{B_s}}

\newcommand{\thspace}{\kern.08333em}

\def \beq{\begin{equation}}
\def \eeq{\end{equation}}
\def \beqn{\begin{eqnarray}}
\def \eeqn{\end{eqnarray}}

\begin{document}
\rightline{TECHNION-PH-97-11}
\rightline{JLAB-TH-97-~~~~~~~~~~~~}
\rightline{BNL-~~~~~~~~~~~~~~~~~~~~~~~}
\bigskip
\bigskip
\bigskip
\centerline{\bf Mixing-induced CP Asymmetries in Radiative $B$ Decays}
\centerline{\bf in and beyond the Standard Model}

\bigskip
\centerline{\it David Atwood$^a$, Michael Gronau$^b$, and Amarjit
Soni$^c$}
\bigskip

\centerline{$^a$Theory Group, CEBAF, Newport News, VA\ \ 23606}
\centerline{$^b$Department of Physics, Technion, Haifa 32000, Israel}
\centerline{$^c$Theory Group, Brookhaven National Laboratory, Upton, NY\ \
11973}
\vskip 4cm
\centerline{\bf ABSTRACT}
\medskip
\begin{quote}
In the Standard Model (SM) the photon in radiative
$\bob$ and $\bsb$ decays is
predominantly left-handed. Thus, mixing induced CP asymmetries in $b\to
s\gamma$
and $b\to d\gamma$ are suppressed by $m_s/m_b$ and $m_d/m_b$, respectively,
and are very small. In many  extensions of the SM, such as the left-right
symmetric model (LRSM), $SU(2)\times U(1)$ models with exotic fermions and
SUSY, the amplitude of right-handed photons grows proportional to the
virtual heavy fermion mass, which can lead to large asymmetries. As an example,
in the LRSM,
asymmetries larger than 50$\%$ are possible even when radiative
decay rate measurements agree with SM predictions.
\end{quote}
\medskip
\leftline{\qquad PACS codes:  11.30.Er, 12.60.Cn, 13.20.Hw}
\newpage

$B$ meson decays may exhibit CP violation effects in a variety of manners
\cite{BCP}. In the Standard Model (SM), $B^0$ decays to CP eigenstates such as
$J/\psi K_S$ involve large time-dependent rate asymmetries between $B^0$ and
$\bar B^0$, which are given in terms of pure CKM phases. On the other hand,
certain asymmetries, such as in $B_s \to J/\psi\phi$, are expected to be
extremely
small in the SM, and are therefore very sensitive to sources of CP violation
beyond the SM. This example represents a class of decay processes, in which
large measurable effects of new physics in CP asymmetries
originate in additional sizable contributions to $\bq-\bqb~(q=d,s)$ mixing
\cite{GL}. Much smaller effects, which are harder to measure
and have considerable theoretical uncertainties, can occur as new
contributions to neutral $B$ decay amplitudes \cite{GWOR}.
Similarly, theoretical calculations of CP violation in charged B decays
entail sizeable uncertaintiues due to final state interaction phases and
therefore, as a rule, cannot be used as unambiguous signals of new physics.

In the present Letter we wish to demonstrate a new way in which large CP
asymmetries in neutral $B$ decays can be introduced by new physics in processes
where the SM predicts very small CP violation. We consider radiative $\bo$ and
$\bs$ decays, $\bo,\bs \to M^0\gamma$, where $M^0$ is any hadronic
self-conjugate
state $M^0 = \rho^0,\omega,\phi,K^{*0}$~(where $K^{*0}\to K_S\pi^0$), etc. As
in $B^0 \to J/\psi K_S$ and $B_s \to J/\psi\phi$, the
asymmetries in $B\to M^0\gamma$ are due to the interference between
mixing and decay. The final states are not pure CP-eigenstates. Rather,
in the SM, they consist to a very good approximation of equal admixtures of
states with positive and negative CP-eigenvalues. Thus, due
to an almost complete
cancellation between contributions from positive and negative CP-eigenstates,
the asymmetries in $b\to q\gamma$ are very small. They are given by $m_q/m_b$,
where the quark masses are current masses. This situation
can be significantly modified in certain models beyond the SM
by new terms in the decay amplitude. This rather special mechanism is to be
contrasted with the more common new physics effect, in which large CP
violation is due to additional contributions to $\bq-\bqb$ mixing.

While our focus will be on {\it mixing-induced} CP violation, let us recall for
completeness that {\it direct} CP violation in radiative $B$ decays, occurring
also in charged $B$ decays, was already studied in the past both for the
exclusive \cite{WYLER}\cite{EILAM}\cite{SONI} and inclusive \cite{SOAR}
case. These effects depend on final state rescattering phases. Since final
states in interesting {\it exclusive} decays involve a hadron and a photon,
electromagnetic (soft) final state phases are small
and can be neglected. The remaining strong phases, originating in the
absorptive
part of the penguin amplitude, can be calculated perturbatively. The
calculation,
which includes bound state effects, involves a fair amount of model-dependence.
The resulting asymmetries in the SM are at a level of 1$\%$ and 10$\%$ for
processes such as $B\to K^*\gamma$ and $B\to\rho\gamma$, respectively
\cite{WYLER}. Asymmetries in {\it inclusive} $b\to s\gamma$ and $b\to d\gamma$,
were calculated in the SM and were found to be at most at this level and
probably smaller \cite{SOAR}. Inclusive asymmetries were also calculated in
models beyond the SM. In a two-Higgs-doublet model containing flavor-changing
neutral Higgs exchange, the asymmetry in $b \to s \gamma$ can reach at most a
level of 10$\%$ \cite{WW}. This would provide some evidence for new physics,
albeit an uncertainty in calculating final state interaction effects.
On the other hand, in the left-right symmetric model
the asymmetries were found to be at most only slightly larger than in the SM
\cite {ARA}, which would be insufficient to signal new physics.

As we will show below, mixing-induced CP asymmetries in exclusive
radiative $\bo$ and $\bs$ decays, both from $b\to s\gamma$ and $b\to d\gamma$,
are very small in the SM and can be 50$\%$ and larger in the
left-right symmetric model, for instance. Such asymmetries would be unambiguous
evidence for physics beyond the SM. The general nature of our argument, which
does not depend on assumptions about final state interactions, will be
explained first.

The processes $b\to q\gamma$ ($q=d,s$) can be described by
the  dipole type effective Lagrangian \cite{INAMI}
\beq
H_{eff} = -\sqrt{8}G_F {e m_b\over 16\pi^2}F_{\mu\nu}
[{1\over 2}F_L^q\  \bar q \sigma^{\mu\nu}(1+\gamma_5)b
+{1\over 2}F_R^q\  \bar q \sigma^{\mu\nu}(1-\gamma_5)b]~.
\eeq
$F_L^q$ is the amplitude for the emission of {\it left} polarized
photons in $b$ (i.e.\ $\bar B$-meson) decay, and $F_L^{q*}$ is the amplitude
for the emission of {\it right} polarized photons in $\bar b$ (i.e.\ $B$-meson)
decay. Similarly, $F_R^q$ is the amplitude
for the emission of {\it right} polarized photons in $b$ decay and $F_R^{q*}$ is
the amplitude for the emission of {\it left} polarized photons in $\bar b$
decay. In the SM
\beq
{F_R^q\over F_L^q} \approx {m_q \over m_b}~,
\eeq
where the masses are current masses. Thus the photons emitted from these $b$
decays are predominantly left-handed. This feature, which is a
key point in our argument, can be easily understood. The term proportional to
$F^q_L$ has the helicity structure $b_R\to q_L\gamma_L$ while the $F^q_R$
term describes
$b_L\to q_R\gamma_R$. In the SM penguin diagram with $W$ exchange,
only the left-handed components of the external fermions couple to the W;
therefore helicity flip must occur on an external leg. Helicity flip on
the $b$-quark leg is proportionl to $m_b$ and contributes to $F^q_L$, while
helicity flip on the $q$-quark leg is proportinal to $m_q$ and contributes to
$F^q_R$. This argument holds to all orders in strong interactions
since the QCD interaction preserves quark helicities.

CP asymmetries in radiative neutral $B$ decays, which follow from the
interference of mixing and decay \cite{BCP}, require that both $B$ and
${\bar B}$ decay to a common state. That is, both should decay to states with
the same photon helicity. (States with different helicities do not interfere
quantum-mechanically, since {\it in principle} the photon helicity can be
measured). Thus, the asymmetry in $b\to q\gamma$ vanishes in the limit
$F^q_R/F^q_L=0$. In the SM these mixing-induced
asymmetries are therefore expected to be quite small, at most of the order of a
few percent in $b\to s\gamma$ and even smaller in $b \to d\gamma$. Furthermore,
the asymmetries depend on the weak phases
associated with $\bq-\bqb$ mixing
and on the phases of the $b\to q\gamma$ decay amplitudes. The
dependence of the asymmetries on time and on the weak phases occuring in
various cases will be discussed below.

Larger CP asymmetries can occur in extensions of the SM
in which the amplitudes of radiative $b$ decays can receive
additional contributions from penguin diagrams with a heavy {\it right-handed}
internal fermion $f$. If a left-to-right
helicity flip occurs on the internal fermion line, then the
amplitude for producing right handed photons will have an additional
enhancement of $m_f/m_b$ with respect to the SM.
There are a number of models with this property, which are potential
candidates for large time-dependent CP asymmetries in radiative $\bo$ and $\bs$
decays. A few examples are the $SU(2)_L\times SU(2)_R\times U(1)$ left-right
symmetric model \cite{MOHAP} to be studied below, $SU(2)\times U(1)$ models
with exotic fermions (mirror or vector-doublet quarks) \cite{LANLON}, and
nonminimal supersymmetric models \cite{MASIERO}. The last two classes of models
will be investigated elsewhere \cite{AGS}.

Let us consider in some detail the time-dependence of a generic exclusive decay
process
\beq
\bq(t) \to M^0\gamma~,
\eeq
for a state which is identified (``tagged'') as a $\bq$ (rather than a $\bqb$)
at time $t=0$.~~$M^0$ is any hadronic self-conjugate state, with CP eigenvalue
$\xi=\pm 1$. We denote
\begin{eqnarray}
A(\bar B\to M^0\gamma_L)&=&A\cos\psi e^{i\phi_L}~, \nonumber\\
A(\bar B\to M^0\gamma_R)&=&A\sin\psi e^{i\phi_R}~, \nonumber\\
A(B\to M^0\gamma_R)&=&\xi A\cos\psi e^{-i\phi_L}~, \nonumber\\
A(B\to M^0\gamma_L)&=&\xi A\sin\psi e^{-i\phi_R}~.
\label{amps}
\end{eqnarray}
For simplicity, we have suppressed the index describing the flavor $q$ of
the neutral $B$ meson and the flavor of $q'$ in the underlying $b\to q' \gamma$
decay. $\psi$ gives the relative amount of left-polarized photons
and right-polarized photons in $\bqb$ decays, $\phi_{L,R}$ are CP-odd weak
phases, while electromagnetic final state phases are absorbed in the amplitude
$A$ (which controls the overall rate) and can be neglected.

Using the time-evolution of a state $\bq(t)$, which oscillates into a mixture
of $\bq$ and $\bqb$ and decays at time $t$ to $M^0\gamma$, we find the
time-dependent decay rate
\beq
\Gamma(t)\equiv\Gamma(\bq(t)\to M^0\gamma)=e^{-\Gamma t}|A|^2[1+\xi
\sin(2\psi)\sin(\phi_M-\phi_L-\phi_R)\sin(\Delta mt)]~.
\eeq
$\phi_M$ is the phase of $\bq-\bqb$ mixing, which is model-dependent.
The corresponding rate $\bar\Gamma(t)$ for an initial $\bqb$ is similar,
however the second term appears with opposite sign. Thus, one finds a CP
asymmetry
\beq
A(t)\equiv {\Gamma(t)-\bar\Gamma(t)\over \Gamma(t)+\bar\Gamma(t)}
=\xi\sin(2\psi)\sin(\phi_M-\phi_L-\phi_R)\sin(\Delta mt)~.
\label{asym}
\eeq
In the above we have neglected, as usual, the small width-difference between
the two neutral $B$ meson states and denoted their mass-difference by
$\Delta m$.
We have also neglected {\it direct} CP violation. As explained in the
introduction, such asymmetries are expected to be small in the SM, at most of
order 1$\%$ and 10$\%$ in (exclusive) $b\to s\gamma$ and $b\to d\gamma$,
respectively. They would show up as additional small $\cos(\Delta mt)$ terms in
the asymmetry \cite{BCP}.

The expression Eq.(\ref{asym}) is similar to the well-known form of an
asymmetry obtained for decays to CP-eigenstates, such as
$\bo\to J/\psi K_S$. The new factor $\sin(2\psi)$ describes helicity
suppression, following from the opposite helicities to which $\bq$ and $\bqb$
prefer to decay. It is the origin of the small asymmetry expected in the SM.

So far, the expression for the asymmetry is general. Now let us consider the
asymmetries within the SM for the four cases of $\bo$ and $\bs$ decays from
$b\to s\gamma$ and $b\to d\gamma$.
In these cases we find
\begin{eqnarray}
{\rm for}~\bo&:&~~~\phi_M=2\beta~, \nonumber\\
{\rm for}~\bs&:&~~~\phi_M=0~,
\label{sm1}
\end{eqnarray}
and
\begin{eqnarray}
{\rm for}~b\to s\gamma : ~~~\sin(2\psi)\approx {2m_s\over m_b}~,~~~\phi_L=\phi_R
\approx 0~,
\nonumber\\
{\rm for}~b\to d\gamma : ~~~\sin(2\psi)\approx {2m_d\over m_b}~,~~~\phi_L=\phi_R
\approx \beta~,
\label{sm2}
\end{eqnarray}
where $-\beta$ is the phase of $V_{td}$ in the standard convention \cite{BCP}.
We note that in $\bo$ and $\bs$ decays to nonstrange states
the asymmetry vanishes identically, due to a cancellation between the weak
phases appearing in $\bq-\bqb$ mixing and in the decay amplitudes. In decays
to strange final states, the asymmetry is proportional to $\sin(2\beta)$. The
sign of the asymmetry is determined also by $\xi$, the CP-eigenvalue of the
hadron $M^0$. We list a few examples of asymmetries expected in the SM:
\begin{eqnarray}
\bo \to K^{*0}\gamma &:& A(t) \approx (2m_s/m_b)\sin(2\beta)\sin(\Delta mt)~,
\nonumber\\
\bo \to \rho^0\gamma &:& A(t) \approx 0~,
\nonumber\\
\bs \to \phi\gamma &:& A(t) \approx 0~,
\nonumber\\
\bs \to K^{*0}\gamma &:& A(t) \approx -(2m_d/m_b)\sin(2\beta)\sin(\Delta mt)~,
\label{examples}
\end{eqnarray}
where $K^{*0}$ is observed through $K^{*0}\to K_S \pi^0$.

Now we turn to the left-right symmetric model (LRSM) \cite{MOHAP}, in order to
study the asymmetries in this extension of the SM. We will limit our analysis to
the most commonly discussed version based on a discrete $L \leftrightarrow R$
symmetry, in which $g_R=g_L$ and the left and right quark mixing matrices are
related to each other either by $V^R=V^L$ or by $V^R=(V^L)^*$. A very strong
lower limit on the
$W_R$ mass was obtained from the $K_L-K_S$ mass-difference \cite{BBS}
\beq
m(W_2) > 1.4~{\rm TeV}~,
\label{wr}
\eeq
and a rather stringent upper bound on $W_L-W_R$ mixing was derived from
semileptonic $d$ and $s$ decays \cite{WOLF}:
\beq
0 \le \zeta \le 3\times 10^{-3}~,
\label{zeta}
\eeq
where \cite{BS1}
\beq
\left( \begin{array}{c}
W_1^+\\
W_2^+
\end{array} \right)
=
\left( \begin{array}{cc}
\cos\zeta & e^{-i\omega}\sin\zeta \\
-\sin\zeta & e^{-i\omega}\cos\zeta
\end{array} \right)
\ \ \
\left( \begin{array}{c}
W_L^+\\
W_R^+
\end{array} \right)~.
\eeq
The limit on $\zeta$ assumes a small CP violation phase $\omega$
and becomes somewhat weaker for large phases \cite{LANG}. In general,
$0 \le \omega \le 2\pi$.

It is generally believed that, due to these rather severe limits, this model
has no important new CP violation effects in the $B$ meson system \cite{GL}.
We will show that, in fact, radiative neutral $B$ decays can have very large
mixing-induced CP asymmetries.
We also note in passing that if the discrete $L \leftrightarrow R$ symmetry is
abandoned, the above constraints on the parameters of the model loosen
substantially \cite{LANG}, in which case it can have sizable nonstandard
effects in $B$ physics \cite{RIGHTB}. However, as
mentioned before, we are going to insist on the discrete symmetry and
consequently on Eqs.(\ref{wr})(\ref{zeta}).

The process $b \to s\gamma$ was studied within the left-right symmetric model
by several authors \cite{LRSM1}\cite{LRSM2}. In addition to the SM
penguin operator with $W$ (and $t$) exchange, the amplitude contains two
penguin-type terms which are potentially large: An amplitude with $W_L-W_R$
mixing and an amplitude involving charged scalar exchange. Both amplitudes
contain an enhancement factor $m_t/m_b$ due to helicity flip on the
internal $t$ quark line. For illustration purposes, we will adopt
the results of Ref.\cite{LRSM2} for the first contribution including QCD
corrections and will neglect the charged Higgs term assuming that the charged
Higgs is sufficiently heavy (e.g. $m_H>20$ TeV). We consider the case $V^R=V^L$.
The terms $F^q_L$ and $F^q_R$ describing the amplitudes for emission of left
and right-handed photons in $b\to q\gamma$ are given approximately by:
\begin{eqnarray}
F_L\propto F(x)+ \eta_{QCD} + \zeta {m_t\over m_b} e^{i\omega}\tilde F(x)~,
\nonumber\\
F_R\propto \zeta {m_t\over m_b} e^{-i\omega}\tilde F(x)~, \label{lrmodel}
\end{eqnarray}
where $x=(m_t/m_{W_1})^2,~\eta_{QCD}=-0.18$,
and the functions $F$ and $\tilde F$ are defined as \cite{LRSM2}
\begin{eqnarray}
F(x) &=&
{x (7-5x-8x^2) \over 24(x-1)^3}-{x^2(2-3x)\over 4 (x-1)^4}\log x~,
\nonumber\\
\tilde F(x) &=& {-20+31x-5x^2\over 12 (x-1)^2}
+{x(2-3x)\over 2 (x-1)^3}\log x~.
\end{eqnarray}
Note that the ratio of the left and right helicity amplitudes for $b\to
q\gamma$
does not depend on the $q$ quark flavor. The two amplitudes are proportional
to a common QCD factor and to equal left and right CKM factors
$V_{tb}{V^*}_{tq}$.

The term $F(x)+\eta_{QCD}$ in (\ref{lrmodel}) is the SM
result, while the terms which involve the small mixing $\zeta$ are
proportional to $m_t/m_b$. For the latter ratio we use a pole mass $m_t=$ 175
GeV, and $m_b(\mu=m_t)=$ 3 GeV, which is obtained from a pole mass of 4.8 GeV.
The $m_t/m_b$ enhancement and the factor $\tilde F(x)/[F(x)+\eta_{QCD}]=2.1$
partially overcome the stringent bound on $\zeta$ in (\ref{zeta}).
Consequently,
as pointed out in Ref.\cite{LRSM2}, the effect of $W_L-W_R$ mixing on the rate
of $b\to q \gamma$ may be significant. Using the above values, we find
the following expression for the LRSM and SM ratio of rates
\beq
{\Gamma(LRSM)\over \Gamma(SM)} \approx |e^{-i\omega} + z|^2 +z^2~,
\label{ratioLRSM}
\eeq
where $z=120\zeta$.

The CP asymmetry Eq.(\ref{asym}) results from an intereference of $F_L$ and
$F_R$, and depends on the two parameters describing $W_L-W_R$ mixing,
the mixing parameter $\zeta$ and the CP violating phase $\omega$. We find the
following expressions for the parameters which determine the
asymmetry in $\bq\to X_{q'} \gamma$ independently of the flavors $q$ and $q'$
\begin{eqnarray}
\tan\psi &\approx& {z\over |e^{-i\omega} + z|}~,\nonumber\\
\phi_L+\phi_R &\approx& {\rm Arg}(e^{-i\omega} + z)~.
\end{eqnarray}
\label{asymLRSM}
The phase of $\bq-\bqb$ mixing is unaffected by new LRSM contributions
\cite{GL},
and is approximately the same as in the SM, $\phi_M=2\beta$ and $\phi_M=0$ for
$\bo$ and $\bs$, respectively.

The parameters $\zeta$ and $\omega$ are constrained by the agreement between
the calculation of the branching ratio for $B\to X_s\gamma$ within the SM
\cite{CMM}, $B(B\to X_s\gamma)_{\small\rm SM} = (3.28\pm0.33)\times 10^{-4}$,
and experiment \cite{CLEO}, $B(B\to X_s\gamma)_{\small\rm EXP} =
(2.32 \pm 0.67)\times 10^{-4}$.
Adding the theoretical and experimental errors in quadrature,
we get,
$B(B\to X_s\gamma)_{\small\rm EXP} / B(B\to X_s\gamma)_{\small\rm SM}=0.71
\pm 0.22$.
The resulting constraint on $\zeta$ and $\omega$,
based on a 90$\%$ confidence level
agreement between theory and experiment (i.e. $.71\pm .36$),
is shown in Fig.1. This figure also illustrates
the large area in the $\zeta, \omega$ plane which would be excluded,
in the future, by a 10$\%$
agreement between the SM prediction and experiment, leaving only two narrow
bands of allowed values around $120\zeta=-\cos\omega$ and above $\zeta=0$.

It is crucial to note that the radiative rate measurements do not probe photon
helicities. Indeed, the
CP asymmetries, which do depend on photon helicities, may be quite
large even when the rate agrees with the SM expectation. In this case we have
\beq
|e^{-i\omega} + z|^2 +z^2 = 1~,
\eeq
a solution of which is $z=-\cos\omega$. Consequently,
$\sin(2\psi) = \vert\sin(2\omega)\vert,~~\phi_L + \phi_R = \pm\pi/ 2$,
where the $+$ and $-$ signs correspond to $0<\omega<\pi$ and $\pi<\omega<2\pi$,
respectively. The asymmetry is given by
\beq
A(t) = \mp\xi\vert\sin(2\omega)\vert\cos\phi_M\sin(\Delta mt)~.
\eeq
The largest asymmetry is obtained when $\zeta$ takes its present experimental
upper limit $\zeta=0.003$, corresponding to $\vert\sin(2\omega)\vert=0.67$.
(The limit on $\zeta$ is actually somewhat higher for $\omega\ne 0$
\cite{LANG}, and the asymmetry can be correspondingly larger).
In this case we find, instead of the SM predictions Eqs.(\ref{examples}),
\begin{eqnarray}
\bo \to K^{*0}\gamma &:& A(t) \approx \mp 0.67\cos(2\beta)\sin(\Delta mt)~,
\nonumber\\
\bo \to \rho^0\gamma &:& A(t) \approx \mp 0.67\cos(2\beta)\sin(\Delta mt)~,
\nonumber\\
\bs \to \phi\gamma &:& A(t) \approx \mp 0.67\sin(\Delta mt)~,
\nonumber\\
\bs \to K^{*0}\gamma &:& A(t) \approx \mp 0.67\sin(\Delta mt)~.
\label{examplesLR}
\end{eqnarray}
All four asymmetries can be larger than 50$\%$. ($10^{\circ}<\beta<35^{\circ}$
\cite{BCP}.)

We comment briefly on expected branching ratios for the interesting processes.
We focus our attention on processes of the type $\bq\to M^0\gamma$, where $M^0$
is a single (unstable) meson state. One may also consider three-body decays,
such as $\bq\to P^+P^-\gamma$ ($P=\pi, K$), however this would require
separating different angular momentum $P^+P^-$ states which have specific
CP-values. So far, only the exclusive $B\to K^*\gamma$
was measured \cite{BK*GAMMA},
$B(B\to K^*\gamma)\sim 4.5\times 10^{-5}$. The corresponding branching ratio of
$B^0\to\rho^0\gamma$ is expected to be lower by a factor $|V_{td}/V_{ts}|^2$
in the SM, which would make it about an order of magnitude smaller.
($0.15<|V_{td}/V_{ts}|<0.33$ \cite{BCP}). One expects $B(\bs\to \phi\gamma)\sim
B(B\to K^*\gamma)$ and $B(\bs\to K^*\gamma)\sim B(B^0\to \rho^0\gamma)$.

Measuring an asymmetry requires tagging and, if carried out in a $e^+e^-$
collider operating at the $\Upsilon(4S)$, it also needs time-dependence.
That is, one must measure the distance of the $B$ decay point away from its
production. In this respect, the $\rho^0$ and $\phi$ can be easily handeled by
their prompt and dominant decays to a pair of pions and kaons, respectively.
On the other hand, due to the $K_S$ finite lifetime, it would be hard to
trace a $K^{*0}$ decaying to $K_S\pi^0$ back to its point of production.
In addition, the $K^{*0}$ decay, via $K_s \pi^0$,
to the final $\pi^+\pi^-\pi^0$ state involves a
suppression factor of 1/9, which makes the effective rate of $\bo\to
K^{*0}\gamma$ comparable to that of $\bo\to\rho^0 \gamma$.
In hadronic colliders, where $B\bar B$ pairs are produced incoherently, no
time-dependence is required.
As we have shown, in models such as the LRSM, the above modes and similar ones
may involve large asymmetries of order 10$\%$ or even 50$\%$. With branching
ratios of a few times $10^{-6}$, asymmetries of order 50$\%$ and 10$\%$ should
be within the reach of planned high-intensity sources of $B$ mesons providing
${\cal O}(10^8)$ and ${\cal O}(10^9)$ $B$'s, respectively. Clearly, improved
bounds on the parameters of the model, $\zeta$ and $\omega$ in the case of LRSM,
would be obtained if no asymmetries were found at this level.

In summary, it was pointed out that the {\it inclusive rate} of radiative $B$
decays is sensitive to certain types of new physics. The photon helicity
measured by {\it mixing-induced CP asymmetries} turns out, on the other hand,
to be sensitive to a different type of effect which appears in
some extensions
of the SM such as the LRSM. Other models will be studied elsewhere \cite{AGS}.
Since the SM predicts very small CP violation, observing sizable asymmetries,
irrespective of their precise values, would be a clear signal of physics beyond
the SM.

\bigskip

M.G. wishes to thank the Brookhaven National Laboratory Theory Group
for a congenial atmosphere in which part of this collaboration was carried out.
This research was supported in part by the U.S.-Israel Binational Science
Foundation under Research Grant Agreement 94-00253/2, by the Israel Science
Foundation, and by the U.S. DOE contracts DC-AC05-84ER40150 (CEBAF) and
DE-AC-76CH00016 (BNL).
\newpage
\noindent
{\bf Figure Caption}
\medskip

\noindent
Fig.1: Presently allowed values of $\zeta$ and $\omega$ from
$B(B\to X_s\gamma)$, deduced by setting EXP/SM $= 0.71\pm 0.36$ (i.e.
to 90$\%$ CL), are included in the shaded area and in the blank internal area.
Only the shaded region would be allowed when a 10$\%$ agreement between the SM
prediction and experiment is attained in the future.
\newpage

\end{document}